\documentclass{iopart}

\usepackage{graphicx}
\usepackage{amssymb}
\begin{document}

\title[Prisoner's Dilemma revisited: evolution of cooperation under pressure]
{Prisoner's Dilemma cellular automata revisited: 
evolution of cooperation under environmental pressure }
\author{J. Alonso$^{1}$, A. Fern\'andez $^{1}$ and H. Fort$^{2}$}
\address{$^{1}$Instituto de F\'isica, Facultad de Ingenier\'ia, 
Universidad de la Rep\'ublica, Julio Herrera y Reissig 565, 11300 Montevideo, Uruguay.\\
$^{2}$Instituto de F\'isica, Facultad de Ciencias, Universidad de 
la Rep\'ublica, Igu\'a 4225, 11400 Montevideo, Uruguay}

\begin{abstract}
We propose an extension of the evolutionary Prisoner's Dilemma cellular 
automata, introduced by Nowak and May \cite{nm92}, in which the pressure 
of the environment is taken into account. 
This is implemented by requiring that individuals need to collect a minimum 
score $U_{min}$, representing indispensable resources (nutrients, energy, 
money, etc.) to prosper in this environment. 
So the agents, instead of evolving just by adopting the behaviour of the 
most successful neighbour (who got $U^{msn}$), also take into account 
if $U^{msn}$ is above or below the threshold $U_{min}$.    
If $U^{msn}<U_{min}$ an individual has a probability of adopting the opposite 
behaviour from the one used by its most successful neighbour. 
This modification allows the evolution of cooperation for 
payoffs for which defection was the rule 
(as it happens, for example, when the sucker's payoff is much worse than the 
punishment for mutual defection).
We also analyse a more sophisticated version of this model in which the selective 
rule is supplemented with a "win-stay, lose-shift" criterion. 
The cluster structure is analyzed and, for this more complex version we found 
power-law scaling for a restricted region in the parameter space. 
\end{abstract}

\maketitle

\section{Introduction}

Cooperation among animals, either within or between
species, is widespread throughout nature \cite{a87}-\cite{ms99}.
This presents a puzzle for Darwinists
since, according to Darwin's theory, the rule among animals should be
competition, not cooperation. 

Attempting to understand the evolution of cooperation,
Maynard Smith and Price \cite{msp73} applied game theory to interactions 
between competing individuals of the same species that use different 
strategies for survival.
They found that in situations like combat, in which each individual must 
decide whether or not to escalate the fight without knowing his opponent's 
decision, the interests of both combatants are best
served if both decide not to escalate the fight.

$2\times2$ games (2 players making a choice between 2 alternatives),
which showed their usefulness in Economics and Social Sciences \cite{axel84}, 
constitute also a basic tool to model the conflict/cooperation 
situations in Biology \cite{ms82}.
Furthermore, the marriage of Game Theory and Darwinian evolution gave rise to 
a new branch of game theory, namely {\it evolutionary game theory} \cite{wei95}.

In particular one of such games is the {\it Prisoner's Dilemma} (PD), now 
well established as a useful tool for studying cooperative interactions  
among self-interested agents. The PD game comes from an experimental setup 
designed by the researchers at the RAND Corporation M. Dresher and M. Flood.
The game refers to an imaginary situation in which two suspects are
arrested near the scene of a crime. The police don't have enough
evidence to convict the pair on the principal charge. 
The two prisoners are held in separate cells and offered a deal:
If one testifies implicating the other in the principal crime will go free,
while the other, if remains silent, will receive 10 years in prison.
If they both testify against each other, each will receive 5 years.
Finally, if they both remain silent, they will both be convicted by a minor
crime and serve one year. What's the rational choice for each prisoner? 
To remain silent (cooperate with your partner) or to confess (not to
cooperate)? The "dilemma" faced by the prisoners is that, whatever the
other does, each is better off confessing than remaining silent. But the
outcome obtained when both confess is worse for each than the outcome they
would have obtained if both had remained silent. This puzzle illustrates
a conflict between individual and group rationality. A group whose members 
pursue rational self-interest may all end up worse off than a group whose 
members act contrary to rational self-interest. 
Formulated in its general form the PD game involves 
two players each confronting two
choices: cooperate (C)  or defect (D) and each makes his choice without
knowing what the other will do.
The possible outcomes for the interaction of both agents    
are: 1) they can both cooperate: (C,C) and get the ''reward" for mutual
cooperation $R$, 2) they can both defect: (D,D) and get the ''punishment"
for mutual defection or 3)
one of them cooperates and the other defects: (C,D); in that case the one 
who played C gets the "sucker's
payoff" $S$ while agent who played D gets the "temptation to defect" $T$.
The following {\em payoff matrix} summarizes the payoffs for {\it
row} actions when confronting with {\it column} actions:

\vspace{-4mm}
 
\begin{center}
 $${\mbox M}=\left(\matrix{(R,R)&(S,T)\cr (T,S)&(P,P) \cr}\right),$$
\end{center}

with the four payoffs obeying the inequalities: 

\begin{equation}
T>R>P>S.
\label{eq:ine1}
\end{equation}
Clearly it pays more to defect: if your opponent defects, and you  
cooperate you will end up with the worst payoff. On the other hand,
even if your opponent cooperates, you should defect because in that case 
your payoff is $T$ which is higher than $R$.
In other words, independently of what the other player does,
defection D yields a higher payoff than cooperation and is the
{\it dominant strategy} for rational agents.
Nevertheless, reasoning that way both agents get $P$ which is worst 
than $R$.
 
A possible way out for this dilemma is to play the game repeatedly. 
In this iterated Prisoner's Dilemma (IPD), in which condition (\ref{eq:ine1}) 
is supplemented with the condition:
\begin{equation}
2R>S+T, 
\label{eq:ine2}
\end{equation}
there are several strategies that outperform the dominant one-shot 'always D' 
strategy and lead to some non-null degree of cooperation.
The tournaments organized by Axelrod \cite{axel84}, \cite{ah81} in the 80\'s 
were very illuminating. He invited researchers from different fields 
to contribute a strategy, in the form of a computer program, to play the 
Prisoner's Dilemma against each other and themselves repeatedly. Each
strategy specified whether to cooperate or defect based on the previous
moves of both the strategy and its opponent. 
The programs were then ranked according to the total
payoff accumulated. The winning program, was also the simplest: 'TIT FOR TAT'
(TFT), which plays C on the first move, and on all subsequent moves copies
the choice of its opponent on the previous move. 
In an ecological approach \cite{a80}, 
the scores from round two were used to calculate the relative frequencies of
the strategies in a hypothetical population. The strategies were then
submitted to each subsequent round in proportion to their cumulative payoff
in the previous round. In the long run, TFT outcompeted its rivals and
went to fixation. 
Axelrod and Hamilton \cite{ah81} used these ecological competition between 
strategies as a basis for their analysis of the evolution of reciprocal altruism. 
This model is applicable in two opposite situations: on the one hand, in the 
case of higher animals, which can distinguish between their various opponents 
in order to reciprocate \cite{s92}, discouraging thus defection. 
On the other hand, in the case of very simple organisms who have only one 
opponent in its lifetime.

Nowak and May \cite{nm92} found another way to escape from the dilemma:
the incorporation of territoriality in evolutionary game theory favours 
cooperation. The authors proposed simple cellular automata (CA) for general 
ecological systems involving indiscriminating organisms who play against 
several opponents (their neighbours). They neglected all strategical 
complexities or memories of past encounters considering {\it unconditional} 
cellular automata  {\it i.e.}  agents using unconditional strategies 
(each cell is either in a C or D state), as opposed to the conditional
ones like TFT, the {\it simpleton} \cite{rc65} or PAVLOV \cite{kk88}
"win-stay, lose-shift", etc.
Cells simply play repeatedly with their neighbours and in the next round or 
generation adopt the state of the most successful cell of their neighbourhood 
(the one that collected the highest score among the cell itself and its 
neighbours).
Coexistence of both states or behaviours were found for a simplified version 
of the PD in which the punishment $P$ is equal to  
the sucker's payoff $S$ \footnote{Indeed this is the frontier between the PD
game an another interesting game, called {\it chicken} by game theorists 
and {\it Hawk-Dove} (H-D) game by evolutionary biologists, in which the 
punishment for mutual defection is the worst payoff 
{\it i.e.} $T>R>S>P$.}, implying then a "weak dilemma" (maximum punishment
{\it i.e.} the minimum possible value of $P$).
Taking $R=1$ and $P=S=0$ allows to parameterise the payoff matrix in
terms of just the parameter $T$.
Szab\'o and T\"oke \cite{st98} slightly modified the Nowak-May (N-M) 
model with the addition 
of randomness: players are chosen to update their states randomly 
by copying the state of one of its neighbours with a probability depending on
the payoff difference.  They measured the fraction of cooperators $c$ for
different values of the temptation to defect $T$ and found a continuous 
transition from $c=1$ to $c=0$ as $T$ increases.
A problem with these simple spatial games is that if $P$ is augmented until 
it becomes non negligible compared with the reward $R$ then 
cooperation disappears and all the individuals end playing D.

An alternative to go beyond weak dilemmas is to consider more sophisticated 
players, with $m$-steps memory and strategies involving conditional probabilities, 
as Lindgren and Nordahl \cite{ln94} did. They considered
payoff matrices parameterised in terms of two parameters, $T/R$ and $P/R$
($S=0$), and found the evolution of cooperation for payoff matrices beyond
"weak" dilemmas.
However, pursuing as much generality as possible without sacrificing 
the simplicity, which is part of the N-M model beauty, in this paper we explore  
a different approach. Our starting point is realising that, interesting as it is, 
the N-M model lacks a fundamental ingredient, namely that of the stress 
exerted by the environment on the individuals.
This is a crucial factor in order to explain the emergence of cooperation
between self-interested individuals even when they are very simple (without
requiring long term memory nor distinguishing "tags" nor access to
sophisticated strategies and, of course, no rational behaviour). 
The basic idea is that individuals need to collect, when playing with their
$z$ neighbours, a payoff above certain threshold $U_{min}$ in order to prosper. 
In an ecosystem $U_{min}$ 
represents the minimal resources (nutrients, energy, etc.) without which
organisms die; in economics it may correspond to some threshold below which 
the business is no longer profitable, etc. Thus, although D players are the 
most successful for $P-S$ large enough, when $U_{min}>z (P-S)$ they
cannot achieve the critical payoff if surrounded by an entire neighbourhood 
of D's and so some of them will be replaced by C players.    
We use a normalized payoff matrix with $R=1$ and $S=0$. Besides the 3
parameters: $U_{min}$, $T$ and $P$ we include a probability $p$ for players 
of adopting the behaviour that is the opposite of the one used by the most
successful neighbour ($msn$). 
In the simplest model version an individual has a probability of 
behaving different from the $msn$ if the score of the $msn$
is below $U_{min}$. 
This simple recipe allows the evolution of cooperation even when the 
punishment $P$ is relatively soft i.e. when the sucker's payoff is much 
worse than the payoff for mutual defection ($P>>S$). Furthermore, it 
gives rise to states of universal cooperation.
We also consider a more sophisticated hybrid model version in which the 
selective rule of copying the behaviour of the $msn$
is supplemented with a "win-stay, lose-shift" criterion. That is, 
individuals also take into account if their own scores $U$
are below or above $U_{min}$ to update their behaviour. 
This version, although more complex, seems well-grounded since it is 
widely known that Pavlovian strategies play a central role in animal
behaviour \cite{ab1}-\cite{ab5}.  
Moreover, the remarkable experiments conducted by 
Milinski and Wedekind \cite{wm96},\cite{mw98} revealed that, humans engaged 
in social dilemma games use by far strategies of the kind of ''win-stay, lose-shift".

For the different model versions, we explore a subspace of the space of 
parameters $\{T,P,U_{min},p\}$ measuring the fraction of cooperators and 
quantities characterizing the resulting cluster structure.

\section{The Basic Model}

The players, which can adopt only two {\it unconditional strategies} or 
behaviours when playing with their neighbours: cooperate (C) or defect (D), 
are thus represented by binary state cells of a two dimensional automaton. 
In this work we restrict ourselves to square 
grids and two types of neighbourhood:
a){\it von Neumann neighbourhood}, consisting of the $z=4$ first neighbouring 
cells of a given cell, and b)
{\it Moore neighbourhood}, formed by the $z=8$ cells surrounding a
given cell.
Typical grid sizes range from $50 \times 50$ to $500 \times 500$.
Periodic boundary conditions are used. 
The score $U$ of a given player is the sum
of all the payoffs it collects against its neighbours. Tables \ref{tab:z4} and
\ref{tab:z8} summarize the different scores for a player depending on 
the number of C cells in its neighbourhood.
The dynamic is synchronous: all the agents update
their states simultaneously at the end of each lattice sweep.

\begin{table}[htp]
\begin{center} 
\begin{tabular}{|c|c|c|c|c|c|}
\hline
& $4C, 0D$ & $3C, 1D$  & $2C, 2D$ & $1C, 3D$ & $0C, 4D$ \\ 
\hline $C$ & $4$ & $3$ & $2$ & $1$ & $0$ \\ 
\hline $D$ & $4T$ & $3T+P$ & $2T+2P$ & $T+3P$ & $4P$ \\  
\hline
\end{tabular}
\vspace{3mm}
\caption{Score $U$ for a player depending on its state C (row 1) or D (row 2)
and the number of C and D agents in its neighbourhood for the $z=4$ case.}
\label{tab:z4}
\end{center}
\end{table}

\vspace{3mm}

\begin{table}[htp]
\begin{center} 
\begin{tabular}{|c|c|c|c|c|c|c|c|c|c|}
\hline
&$8C, 0D$&$7C, 1D$&$6C, 2D$&$5C, 3D$&$4C, 4D$&$3C, 5D$&$2C, 6D$&$1C, 7D$&$0C, 8D$\\ 
\hline $C$ & $8$ & $7$ &$6$ & $5$ &$4$ & $3$ & $2$ & $1$ & $0$ \\ 
\hline $D$ &$8T$&$7T+P$&$6T+2P$&$5T+3P$&$4T+
4P$&$3T+5P$&$2T+6P$&$T+7P$&$8P$\\  
\hline
\end{tabular}
\vspace{3mm}
\caption{The same as Table \ref{tab:z4} but for $z=8$ neighbours.}
\label{tab:z8}
\end{center}
\end{table}

In the CA of ref. \cite{nm92} natural selection is implemented 
very simply: each agent or player adopts the state of the most
successful neighbour ($msn$) who got $U^{msn}$. 
Here, if $U^{msn}$ is below the threshold $U_{min}$, we allow the individuals 
to adopt the opposite state with a probability $p$.     
The rationale for this is that blindly copying the most
successful neighbour, when its score doesn't reach a critical threshold,
may not be, in the long run, the most efficient strategy from an evolutionary 
point of view.

We consider two possible variants:

\begin{enumerate}

\item {\it Simplest variant: Conditional copying the most successful 
neighbour.}

If $U^{msn}\geq U_{min}$, the player copies the state of its $msn$.
Otherwise, if $U^{msn} < U_{min}$, the player has a 
probability $p$ of adopting the opposite state. 
\\

\item {\it Variant 1 + death of organisms.}

This variant contemplates the possibility that organisms that don't reach 
the threshold $U_{min}$ die and some cells remain unoccupied. 
The rules are the same as above except that in the case when 
$U^{msn}< U_{min}$, instead of adopting the strategy of the $msn$, the
player dies with probability $1-p$ leaving an empty cell. 
An empty cell updates its state copying the one of its $msn$ with 
probability $1-p$ (adopting the opposite behaviour with probability $p$). 
Finally, an 
empty cell surrounded by empty cells remains unoccupied in the next round.

\end{enumerate}

It turns out that, for both variants of the model, the system 
reaches a steady state with a definite value $c$ for the fraction of agents
playing C after a transient. The duration of the transient
depends on the lattice size and the neighbourhood. For instance for 
a $50\times50$ lattice and $z=8$ it last typically between 100 and 200 rounds.

To avoid dependence on the initial conditions, the measures correspond to
averages over an ensemble of 100 systems with arbitrary initial 
conditions. The standard deviation is about 7 percent so this averages are 
quite representative. 

Here, we present results for a subspace of the parameter 
space $\{T,P,U_{min},p\}$. We
choose definite values for the punishment $P$ and the probability 
parameter $p$, specifically: $P=0.5$\footnote{
This value of the punishment implies a non weak 
dilemma and, both for $z=4$ or $z=8$, leads to $c=0$ when simulating
the model of ref. \cite{nm92}.} and $p=0.1$. 
The temptation parameter $T$ is varied between 1 and 2.
A threshold $U_{min} > zP$ is required in order to avoid the all D state; 
on the other hand $U_{min} > zT$ 
doesn't make sense since no one can reach this threshold. 
Thus, the parameter space reduces to the square plane $T-U_{min}$ 
delimited by $1 \le T \le 2$ and  $zP \le U_{min} \le zT$.

After a transient, the steady or asymptotic fraction of C agents $c$
is computed for a grid of points in the $T$-$U_{min}$ plane 
using lattices of relatively modest size: $50\times50$. 
Similar results hold for $100\times100$ lattices or bigger.

\begin{figure}[htp]
   \begin{center}
     \includegraphics[height=8.0cm]{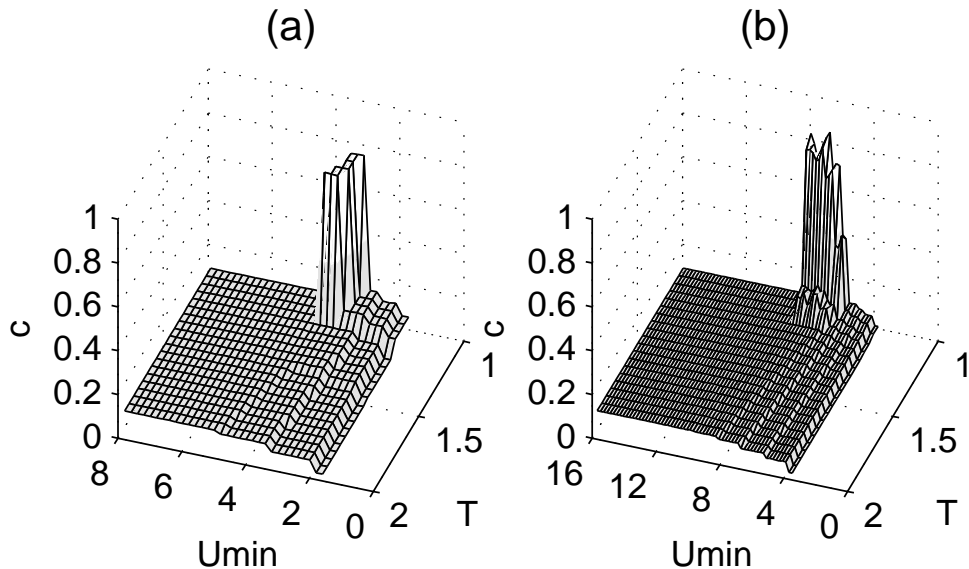}
   \end{center}
\caption{Asymptotic frequency of cooperators for the 
simplest model, for $p=0.1$, (a) z=4 neighbours and (b) z=8 neighbours.}
   \label{fig:simple}
\end{figure}

Figures \ref{fig:simple}.a and \ref{fig:simple}.b 
corresponding to the first model variant show a similar dependence on frequency 
of cooperators with $T$ and $U_{min}$. 
Note that, when $U_{min} > zP$, the fraction of cooperators raises 
from zero to a non negligible value regardless of the value of $T$ 
\footnote{At least in the considered $T$ interval: $1<T<2$.}.
The explanation of this is simple: a D agent surrounded by D's  
gets a score $zP$ that is below the surviving threshold, and thus 
has a probability $p$ of becoming C in the next round. 
Basically three regions can be distinguished in the plots:

\begin{itemize}

\item A {\it stepladder region} emerges from  
the right border $U_{min}=zP$.

\item For not too large values of $T$ and $U_{min}$ there is a 
high {\it peak} of cooperation, delimited to the left by $U_{min}$ = $zR=z$
(when all the cells play C). 

\item Finally, beyond $U_{min}=zR=z$,  $c$ reaches a {\it plateau}  
delimited by the straight line $U_{min}(T)=zT$ ($U_{min}$ greater than
$zT$ is an unreachable score in the game we are considering). 

\end{itemize}

To understand the peak of cooperation it is illuminating to 
consider a small deviation of $T$ from 1: $T =1+\epsilon$. 
Therefore, for $z=8$ and $P=0.5$ the Table \ref{tab:z8} 
becomes the Table \ref{tab:z8epsilon}. For $U_{min}$ greater than 6, 
the only D agents which can achieve the minimum $U_{min}$ are the ones surrounded 
by at least 4 C's (see row 2 of Table \ref{tab:z8epsilon}), so cooperation grows 
dramatically. This corresponds to $\epsilon\lesssim0.16$ (for increasing values 
of $\epsilon$, D agents surrounded by less than 4 C's will survive and we 
cannot expect great values for $c$). When $U_{min}=8$ $c$ drops abruptly since 
even C agents surrounded entirely by other C's cannot survive anymore.

\begin{table}[htp]
   \begin{center} 
      \begin{tabular}{|c|c|c|c|c|c|c|c|c|c|}
     \hline
      & $8C, 0D$ & $7C, 1D$ &  $6C, 2D$ & $5C, 3D$ & $4C, 4D$ & $3C, 5D$  & $2C,6D$ 
      & $1C, 7D$ & $0C, 8D$ \\
      \hline $C$ & $8$ & $7$ &$6$ & $5$ &$4$ & $3$ & $2$ & $1$ & $0$ \\
      \hline $D$ & $8.0+8\epsilon$ & $7.5+7\epsilon$ & $7.0+6\epsilon$ & $6.5+5\epsilon$& 
      $ 6.0+4\epsilon$ & $5.5+3\epsilon$ & $5+2\epsilon$ & $4.5+\epsilon$ & $4$ \\ 
      \hline
      \end{tabular}
   \end{center}
   \vspace{3mm}
   \caption{Same as Table 2 for $T=1+\epsilon$, $P=0.5$}
   \label{tab:z8epsilon}
\end{table}
 
The stepladder structure for $z=8$ (a similar analysis holds for $z=4$) 
can be easily explained considering the scores for D agents of Table 
\ref{tab:z8}. As long as 
$U_{min}$ increases each D agent needs more C agents in its surroundings 
in order to achieve the threshold. So cooperation grows with $U_{min}$ by 
steps at the values mentioned before: $U_{min}=T+7P$, $U_{min}=2T+6P$ 
and so on, which correspond to straight lines with different slopes in 
the $(T,U_{min})$ plane. 
Finally, when $U_{min}>8T$ the minimum required is above any agent's 
possible score, then the fraction of agents C one time step further will be 
given by 
\begin{equation}
c(t+1)=pf_{D}+(1-p)f_{C},
\label{eq:cfC1}
\end{equation} 
where $f_{D}$ stands for the fraction of agents (C and D) whose most 
successful neighbour is a D-cell and $f_{C}$ is the fraction of agents (C and D) 
whose most successful neighbour is a C-cell. As none of the agents achieves the 
threshold, the state of all of them is updated with probability $p$ to a 
state opposite to the one of the $msn$. 
For small values of $p$, $f_{C}\approx0$ (since a C agent needs to be
surrounded by a minimum number of C agents to be the most successful), 
$f_{D}\approx1$ and finally $c\approx p$. 
This explains why the height of the plateau coincides with the probability 
$p$ \footnote{Besides $p=0.1$, we checked this also for $p=0.2$ and
$p=0.3$}.\\

\begin{figure}[htp]
   \begin{center}  
     \includegraphics[height=8.0cm]{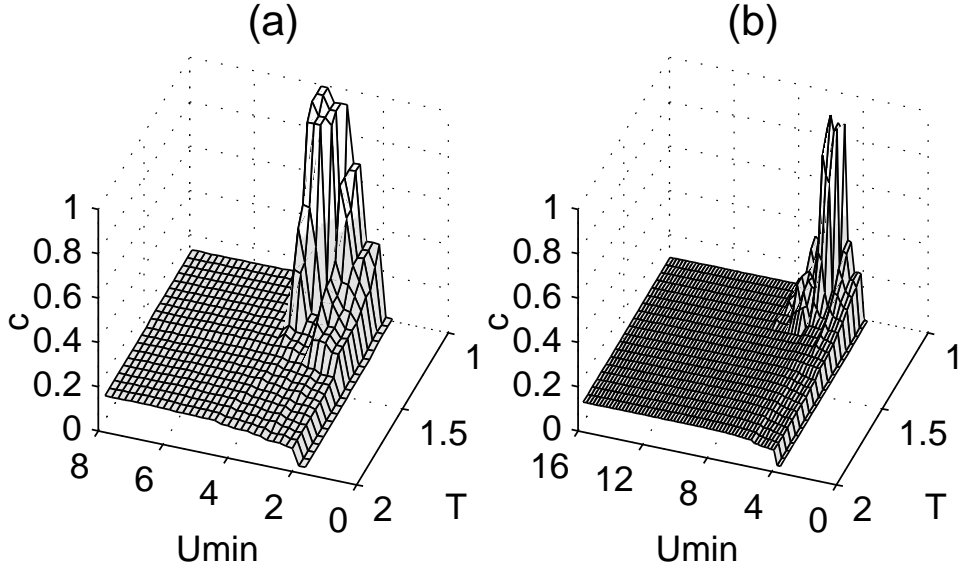}
   \end{center}
   \caption{Asymptotic frequency of cooperators for the empty cells variant,
      for $p=0.1$, (a) z=4 neighbours and (b) z=8 neighbours; frequency is 
normalized by the total number of cells (including empty ones)} 
\label{fig:empty}
\end{figure}

The landscape that emerges from the second model variant 
(see Fig. \ref{fig:empty}.a and Fig. \ref{fig:empty}.b) is very similar to
the one produced by the first variant.
\\

\section{Hybrid Model: Natural Selection Complemented with a Pavlovian Criterion.}

A relevant input for an agent to assess its performance in a game is 
the comparison of its own score $U$ with $U_{min}$.   
If it is above $U_{min}$ then the agent's behaviour may be worth   
keeping even if it is not the most successful in the neighbourhood.
The behaviour updating rule thus becomes more sophisticated and instead 
of the previous rule we have:

\begin{enumerate}

\item  {If $U^{msn}<U_{min}$:}
The player adopts the opposite state of its $msn$ but {\it now} with probability 
$1-p$.

\item  {If $U^{msn}\geq U_{min}$:}
There are two alternatives depending on its score $U$:

\begin{itemize}

\item If $U < U_{min}$, the player keeps its state with probability $p$ 
(adopts the state of its $msn$ with probability $1-p$).

\item If $U\geq U_{min}$ the player keeps its state with probability $1-p$ 
(adopts the state of its $msn$ with probability $p$).

\end{itemize}
\end{enumerate}

Therefore, this model interpolates between the ordinary evolutionary
recipe of copying the most successful neighbour and
the "win-stay, lose-shift" criterion of the game analysed in detail 
by Herz \cite{h94}.

\subsection{Frequency of cooperators}

\begin{figure}[htp]
   \begin{center}  
     \includegraphics[height=8.0cm]{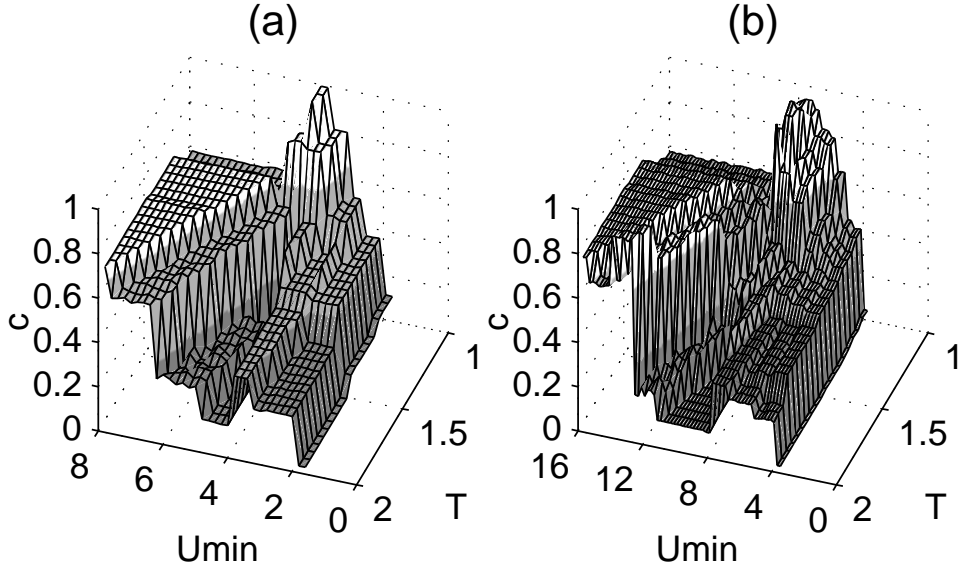}
   \end{center}
   \caption{Asymptotic frequency of cooperators for the hybrid variant, for
   $p=0.1$, (a) z=4 neighbours and (b) z=8 neighbours.}
   \label{fig:hybrid}
\end{figure}

Again, the system reaches a steady state with a definite value $c$ for 
the fraction of agents playing C after a transient.
This hybrid model produces qualitatively similar results, but there 
are remarkable modifications in the landscape 
shown in Fig. \ref{fig:hybrid}. 

\begin{figure}[htp]
   \begin{center}  
     \includegraphics[height=6.0cm]{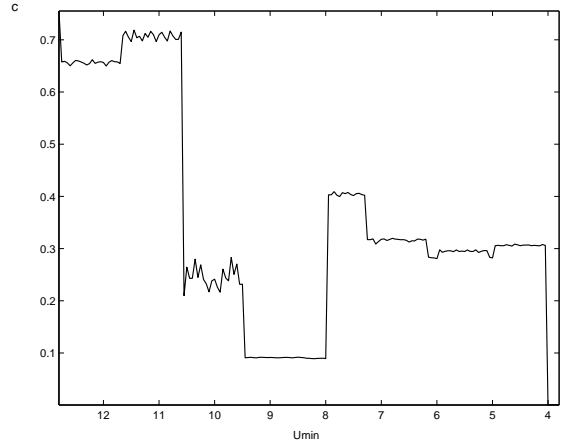}
   \end{center}
   \caption{Asymptotic frequency of cooperators as a function of $U_{min}$ for $T$=1.6, z=8.}
   \label{fig:cvsUm}
\end{figure}

Firstly, we observe a strong 
increase in $c$ for all the parameter space surrounding the peak zone.
In particular, note the height of the plateau and the step formation.
Secondly, most part of the plateau is replaced by steeply "cliffs".
Figure \ref{fig:cvsUm} shows the fraction of cooperators for $T=1.6$, $z=8$.
Between $zP(=4)$ and $zR(=8)$, the fraction $c$ increases almost monotonously with $U_{min}$. However, once $U_{min}$=8 is reached, the fraction of cooperators falls down drastically. This is due to the fact that C agents surrounded entirely by C's can no longer survive and turn into D agents with probability $1-p=0.9$. Since some of the D agents still have a payoff above $U_{min}$, they survive to the next round, so cooperation fraction should decrease. As long as $U_{min}$ increases, more C's are neccesary in the neighbourhood of a D for this agent to keep its strategy, so cooperation would increase again.

When $U_{min}>8T$ the equation (\ref{eq:cfC1}) is replaced by

\begin{equation}
c(t+1)=(1-p)f_{D}+pc(t).
\label{eq:cfC2}  
\end{equation} 
Hence, in the steady state ($c(t+1)=c(t)$) we have the solution $c=f_{D}$. 

\subsection{Cluster structure}

\begin{figure}[htp]
   \begin{center}
     \includegraphics[height=10.0cm]{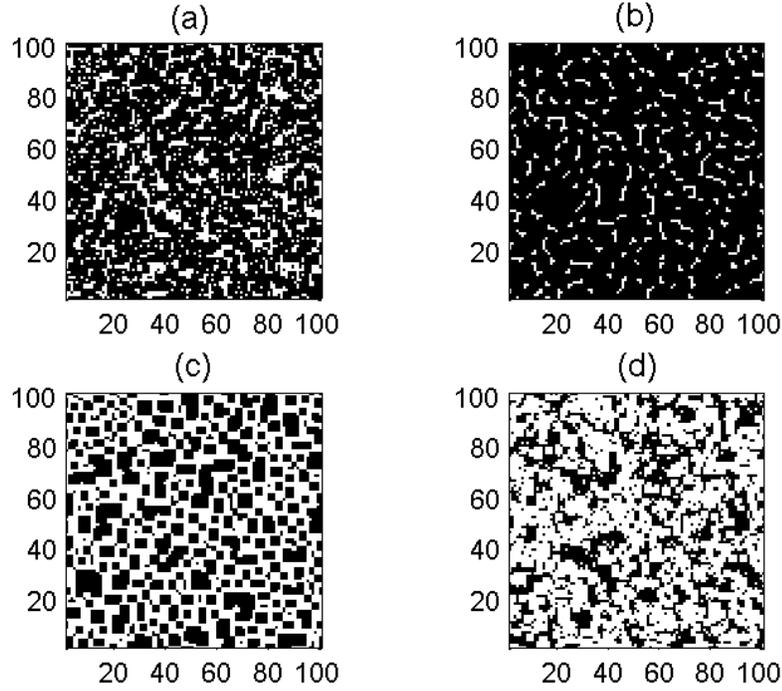}
   \end{center}
	\caption{Steady state cooperation maps for the $z=8$ 
hybrid model (with $p$=0.1) for: 
(a)$T=1.5$, $U_{min}=11.9$ (b)$T=1.06$, $U_{min}=6.9$ (c)$T=1.2$, $U_{min}=5.5$ 
(d)$T=1.6$, $U_{min}=7.5$. Black cells correspond to C agents and white cells to 
D agents. The mean frequency of cooperators corresponding to each map is 
(a)$c=0.75$ (b)$c=0.91$ (c)$c=0.50$ (d)$c=0.40$.}
        \label{fig:lattice}
\end{figure}

An additional novelty of this model, connected with the greater richness in the 
$c$ landscape (see Fig. 3), is the cluster structure which exhibits power law 
scaling for a restricted region in the $T$-$U_{min}$ plane (for the basic model 
no power-laws were found).
In this subsection we analyse the cluster structure and spatial patterns in
the three different regions of the plane $T-U_{min}$ identified in the
previous subsection. 
We present results for the $z=8$ Moore neighbourhood, since it is the one that
exhibits more clear cut results.
Fig. \ref{fig:lattice} shows snapshots of the steady state at four representative
points in the $T-U_{min}$ plane:
(a) [$T$=1.5,$U_{min}$=11.9] belonging to the plateau ($c\simeq$0.75),
(b) [$T$=1.06,$U_{min}$=6.9] belonging to the peak ($c\simeq$0.91),   
(c) [$T$=1.2,$U_{min}$=5.5] at the side of the peak ($c\simeq$0.5) and
(d) [$T$=1.6,$U_{min}$=7.5] belonging to the stepladder region
($c\simeq$0.4).

\begin{figure}[htp]
   \begin{center}
     \includegraphics[height=10.0cm]{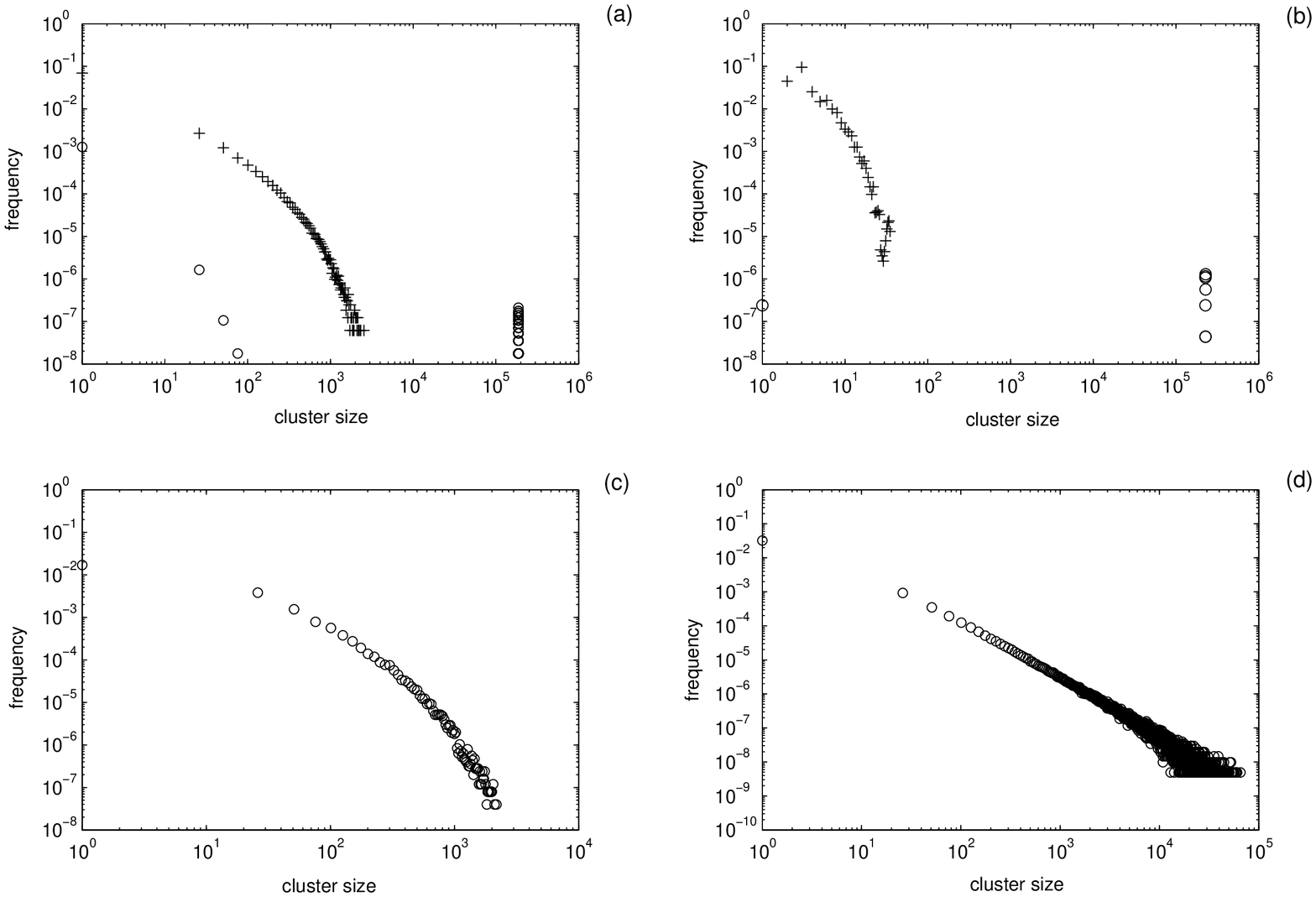}
   \end{center}
	\caption{Dependence of the number of C(o) clusters with size (in logarithmic scale) 
                 for the hybrid model with $z=8$ for (a)$T=1.5$, $U_{min}=11.9$ 
                 (b)$T=1.06$, $U_{min}=6.9$ (c)$T=1.2$, $U_{min}=5.5$ (d)$T=1.6$, 
                 $U_{min}=7.5$ and $p=0.1$. For (a) and (b) (where big C clusters dominate) the dependence of the number of D(+) clusters with size is also shown. 
                 Measures were performed on a $500\times500$ lattice and 
                 clusters sampled over 100 rounds after transient.}
        \label{fig:loglog}
\end{figure}

For [$T$=1.5,$U_{min}$=11.9], although fraction of cooperators is stable the 
spatial patterns change constantly as a consequence of the transition rules. 
One of these patterns is shown in Fig. \ref{fig:lattice}.a.
For [$T$=1.06,$U_{min}$=6.9], giant stable clusters dominate the lattice 
as expected from 
the high level of cooperation in that region as shown in Fig. \ref{fig:lattice}.b.
At the side of the cooperation peak there are spatial stable structures 
of clusters as the ones shown in Fig. \ref{fig:lattice}.c.
When we move away from the peak into the region bounded between $2T+6P$,$4T+4P$ and 
$U_{min}=8$ scale invariance 
emerges: clusters of all size occur as can be seen from Fig. 
\ref{fig:lattice}.d for $T=1.6$ and $U_{min}=7.5$. In this case  
we are in presence of constantly changing spatial patterns again. 
Histograms of the size distribution of clusters for the four above points 
in the $T-U_{min}$ plane are shown in Fig. \ref{fig:loglog}.
The histogram \ref{fig:loglog}.b shows that only exist (giant) clusters for 
a narrow interval of sizes. As we move to regions in the phase space with 
lower values of $c$, a greater diversity of sizes occur (see Figs. \ref{fig:loglog}.a 
and \ref{fig:loglog}.c) until \ref{fig:loglog}.d clearly shows a power law 
distribution with exponent $-1.6357\pm0.0001$. Power laws are the signature of  
organization into a critical state. It indicates that the system exhibits 
the highest pattern of diversity: there are few large structures and many 
smaller clusters.
This power-law scaling emerges only for a very reduced region in the
plane $T-U_{min}$ in the vicinity of the point $[T=1.6,
U_{min}=7.5]$. 
In that sense this scale-free behaviour seems more to ordinary critical 
phenomena (second order phase transitions), where a fine-tuning of the 
control parameters is required, than 
to the much more robust self-organized criticality (SOC). 
\\

\begin{figure}[htp]
   \begin{center}  
     \includegraphics[height=6.5cm]{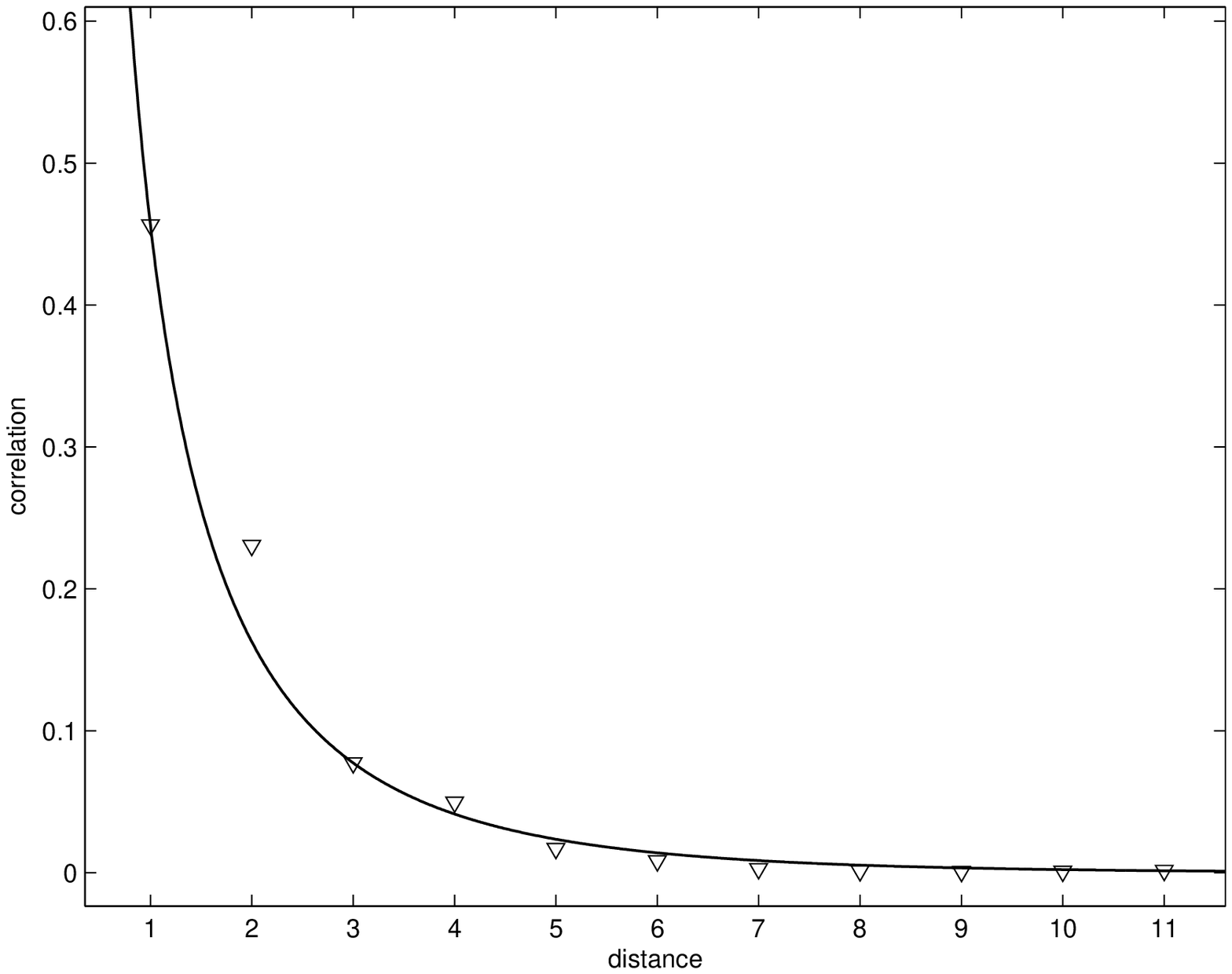}
   \end{center}
   \caption{Correlation between C agents as a function of distance for $[T=1.6,U_{min}=7.5]$; values obtained in simulations ($\triangledown$) and theoretical fitting  $\alpha\frac{e^{-\frac{r}{\xi}}}{r}$ (where $\xi=2.95$) in solid line. }
   \label{fig:corr}
\end{figure}

We also explored the correlation function in this region and found a scaling with distance $r$ proportional to $\frac{e^{-\frac{r}{\xi}}}{r}$. Fig. \ref{fig:corr} shows this function for the point $[T=1.6,U_{min}=7.5]$ where the correlation length $\xi$ takes the value $2.95$.
\\

\begin{figure}[htp]
\begin{center}
\includegraphics[height=10.0cm]{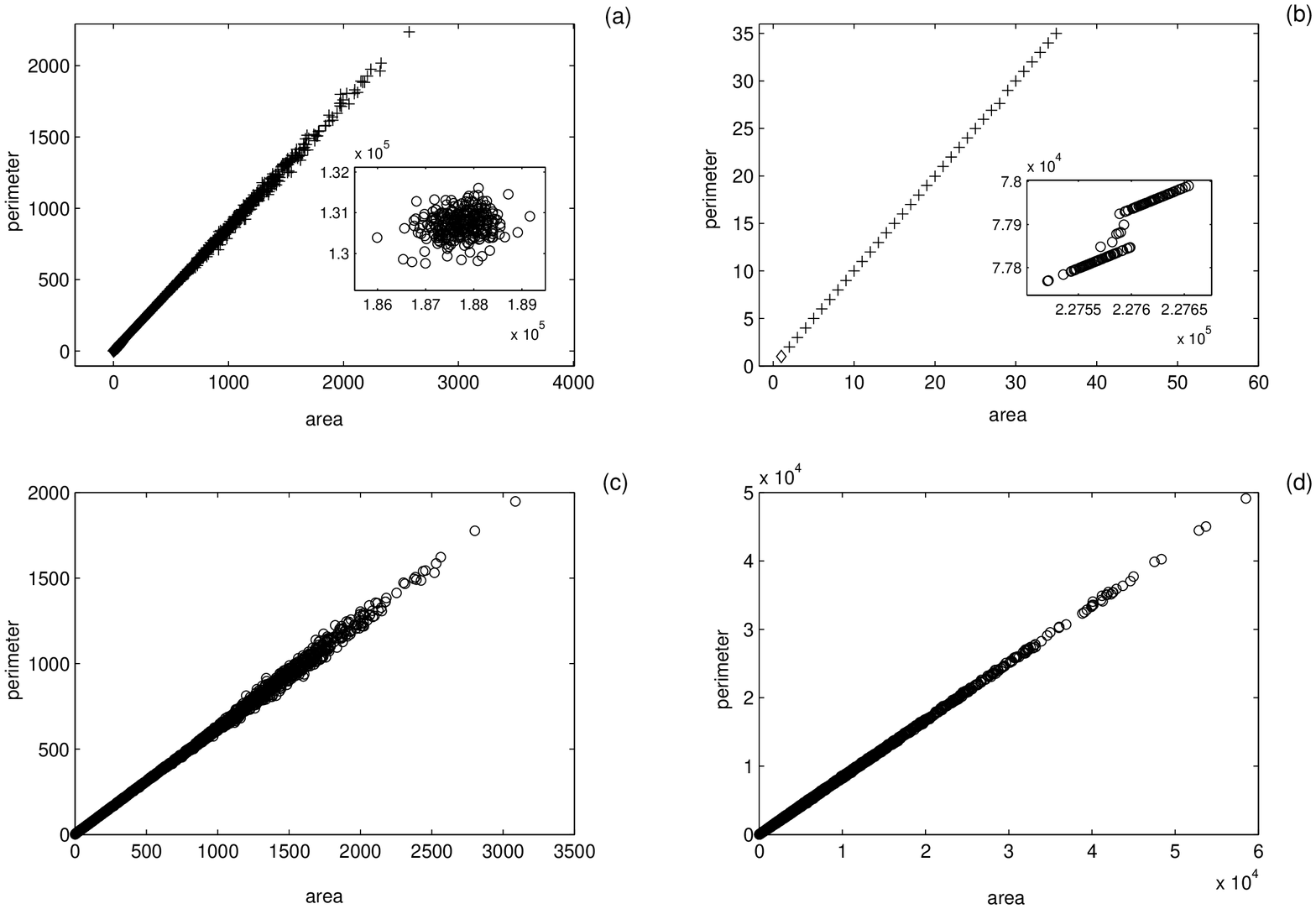}
\end{center}
\caption{The perimeter $\ell$ plotted as a function of C's(o) cluster area $A$
for the hybrid model with $z=8$ for (a)$T=1.5$, $U_{min}=11.9$ 
(b)$T=1.06$, $U_{min}=6.9$ (c)$T=1.2$, $U_{min}=5.5$ (d)$T=1.6$, 
$U_{min}=7.5$ and $p=0.1$. For (a) and (b) perimeter of D's(+) is also shown.
Measures were performed on a $500\times500$ lattice and
clusters sampled over 100 rounds after transient.}
\label{fig:perarea}
\end{figure}

Besides the size distribution of clusters, the relationship between
the perimeter and the area of the clusters provides useful information
on their geometry. 
The area $A$ of a cluster is the number of all connected cells with a
given strategy (C or D) and its perimeter $\ell$ is defined as the 
number of cells that form its boundary (those cells of the cluster 
with at least one neighbour not belonging to it). We compute the 
mean perimeter $\ell(A)$ for a given area $A$ 
averaging over all the perimeters of clusters with given area $A$.
Plots of $\ell$ vs. $A$ for the four $T,U_{min}$ points treated before 
are shown in Fig. \ref{fig:perarea}.  

Figure \ref{fig:perarea}.b is consistent with the 
very narrow range of sizes of C clusters observed in Fig. \ref{fig:loglog}.b (something similar 
happens for Fig. \ref{fig:loglog}.a but without a clear dependence of perimeter with area).     
In Fig. \ref{fig:perarea}.d -which corresponds 
to a power law in size distribution as shown in Fig. \ref{fig:loglog}.d-
the mean perimeter scales linearly with the area.
From this linearity it follows that the ratio of perimeter to              
interior (being the interior $A-\ell(A)$) becomes independent of the cluster size.
The coefficient of the line $\ell(A)$ is $0.8369\pm0.0001$. 
This result differs greatly from the square root dependence of $\ell$ with 
$A$ expected for regular geometry and is an indicator of the ramified 
structure of clusters (see Fig. \ref{fig:lattice}.d).

\section{Discussion}

We have shown how cooperation among self-interested individuals can 
emerge from evolution in PD games, involving quite arbitrary payoff 
matrices (instead of just weak dilemmas), 
using the simplest possible agents: unconditional strategists, 
without long term memory and without distinguishing "tags".
This allows the applicability of the model to a wide variety of contexts
from natural to social sciences.

The main idea was to include the influence of the environment exerting 
pressure on individuals to cooperate even when the
punishment for defecting is relatively soft. This is implemented by requiring 
a minimum score $U_{min}$ necessary for agents to carry on vital
functions. 
In particular, for moderate values of the temptation to defect $T$,
there is an intermediate range of values of $U_{min}$ that 
maximizes cooperation among self-interested agents producing a state of  
"universal cooperation".

Our findings might be connected with questions in evolutionary genetics
like the effects of deleterious mutations on fitness. 
Mutations, in spite of being the ultimate engine for evolution, in general
have a negative effect on fitness. It has been widely accepted that these
deleterious fitness effects are, on average, magnified in stressful    
environments. Recent experimental measures of growth rates of E.coli   
mutants under a diverse set of environmental stresses suggest just the 
opposite: the effects of deleterious mutations can sometimes be ameliorated
in stressful environments \cite{kl03}.
A possibility is that C-inclined organisms may be regarded as deleterious 
mutants in the case of no stress which can take over the population under 
some (appropriate) degree of stress.

It is worth remaking that the more sophisticated model, that results when 
supplementing the ordinary evolutionary
recipe of copying the most successful neighbour with a Pavlovian 
"win-stay, lose-shift" criterion, exhibits two relevant properties.
The first is global optimisation {\it i.e.} it enhances 
the cooperation level. 
The second is the emergence of power-law scaling
in the size distribution for clusters of C-agents. 
Power-laws were also found in a different study of cellular 
automata playing the PD game with Pavlovian strategies \cite{fv05}.
However, in that case, this scaling behaviour is much more robust 
than the one we found here which holds only for quite reduced region in the 
$T-U_{min}$ plane.

The effect of requiring a threshold has been analysed in social sciences. 
For instance, 
in relation to reinforcement learning by Borgers and Sarin \cite{bs97},\cite{bs00},
although treatment is quite different. 
Indeed, this threshold represents an aspiration level and may
improve the decision maker's long-run performance.   

To conclude, we envisage some future extensions of this (these) model (models). 
For instance to explore the effect of heterogeneities, in the environment 
(a landscape dependent $U_{min}$ function) or in the agents 
(different payoff matrices, different types of individuals, etc.).  
In addition, the spatial networks observed in nature are in general 
not uniform square lattices like the ones considered here.
So another interesting direction that seems worth studying is to  
consider more realistic network topologies, for example scale free
\cite{ba99} or small worlds networks \cite{ws98}.    

\ack This work was partially supported by CSIC.

\end{document}